\documentclass[12pt]{article}
\usepackage{graphicx}
\usepackage{fullpage}
\begin{document}
\title{
Film Growth and Surface Roughness with\\
Fluctuating Covalent Bonds 
in Evaporating Aqueous Solution of
Reactive Hydrophobic and Polar Groups:\\
A Computer Simulation Model
}
\maketitle

\begin{center}

Shihai Yang, Adam Seyfarth, Sam Bateman, Ras B. Pandey

Department of Physics and Astronomy

The University of Southern Mississippi

Hattiesburg, MS 39406
\end{center}
\bigskip

\begin{abstract}
A computer simulation model is proposed to study film growth and surface roughness
in aqueous ($A$) solution of hydrophobic ($H$) and hydrophilic ($P$) groups 
on a simple three dimensional lattice of size $L_x \times L_y \times L_z$ with an 
adsorbing substrate. Each group is represented by a particle with appropriate 
characteristics occupying a unit cube (i.e., eight sites). The Metropolis algorithm 
is used to move each particle stochastically. The aqueous constituents are allowed 
to evaporate while the concentration of $H$ and $P$ is constant. 
Reactions proceed from the substrate and bonded particles can hop within a 
fluctuating bond length.
The film thickness ($h$) and its interface width ($W$) are examined
for hard-core and interacting particles for a range of temperature 
($T$). Simulation data show a rapid increase in $h$ and $W$ is followed by 
its non-monotonic growth and decay before reaching steady-state equilibrium 
($h_s, W_s$) in asymptotic time step limit. The growth 
can be described by power-laws, e.g., $h \propto t^{\gamma}, \ W \propto t^{\beta}$
with a typical value of $\gamma \approx 2, \ \beta \approx 1$ in initial time
regime followed by $\gamma \approx 1.5, \ \beta \approx 0.8$ at $T = 0.5$.
For hard-core system, the equilibrium film thickness ($h_s$) and surface roughness 
($w_s$) seem to scale linearly with the temperature, i.e.,
$h_s = 6.206 + 0.302 T, \ W_s = 1,255 + 0.425 T$ at low $T$ and 
$h_s = 6.54 + 0.198 T, \ W_s = 1.808 + 0.202 T$ at higher $T$. 
For interacting functional groups in contrast, $h_s$ and $W_s$ 
decay rapidly followed by a slow increase on raising
the temperature. 
\end{abstract}

\section{Introduction}
Film formation of waterborne two component polyurethanes (WB 2K-PUR) consists of a 
hydrophobic polyisocyanate crosslinker and a hydrophilic polyester polyol in 
aqueous solution \cite{otts1}$-$\cite{march}. Simultaneous progression of 
physicochemical processes such as 
water evaporation, crosslinking reactions, droplet coalescence, etc. leads to 
complex heterogeneous structures. Numerous experimental studies are focused on 
studying the morphology of film surface and its roughness as a function of 
temperature, polymer chain architecture, solvent evaporation, and molecular weight 
etc. \cite{strawhecker, ho},\cite{muller}$-$\cite{lestage2}. The hydrolysis 
reactions 
between isocyanate groups and water leads to formation of polyurea (PUA) while the 
reaction between isocyanate groups and hydroxyl groups of polyol results in 
polyurethane (PUR) \cite{march}. Thus, the water solvent plays an important 
role in orchestrating the evolution of the film growth. Stochastic mobility of 
constituents drives the mixture towards its thermodynamic equilibrium while the 
kinetic reactions may arrest into heterogeneous domains of PUR and PUA. Water loss 
due to evaporation enhances the complexities as the thermal equilibration competes 
with the crosslinking. Recent laboratory data from ATR-FTIR spectra
\cite{march, otts2} provides 
estimates for energy spectrum involved in stretching and bending vibrations of 
various bonds. From the spectral intensity, it is possible to extract the 
concentration of PUA and PUR and gain insight into its dependence on humidity 
(water concentration) \cite{wicks}. AFM images provide information about their 
spatial distributions. In general, experiments show that increasing humidity leads 
to higher roughness \cite{otts1, zhao2}.

A systematic understanding of the global characteristics of the film growth from 
its basic constituents ($H, P, A$) is highly desirable but not feasible with current 
experimental tools. Due to complexity, it is also not feasible to incorporate the 
thermodynamic equilibration involving stochastic mobility of constituents and 
kinetic reactions with covalent bonding in a selfconsistent analytical theory
\cite{wool}. Therefore, computer simulations \cite{tsige}$-$\cite{otts3} remain 
the primary tool to complement and understand the laboratory observations
\cite{otts1}. Some attempts have been made recently to study the film growth 
in such a multicomponent ($H, P$, and $A$) system via computer simulations 
\cite{pandey, otts3}. The constituents ($H, P$, and $A$) are represented by 
particles with appropriate characteristics such as molecular weights, reaction 
functionality, and phenomenological interactions in such coarse grained model on 
a discrete lattice.  The mixture equilibrates as the particles execute their 
stochastic motion via Metropolis algorithm and water component evaporates from the 
top of the sample. The kinetic reactions proceed from the adsorbing substrate and 
the film develops as the reacting particles tether together by covalent bonds. 
Evolution of the density profiles, film thickness, interface width (a measure of 
roughness) are studied as a function of temperature and concentrations of its 
constituents with a range of reaction rates \cite{otts3}. The roughness of the 
film is found to increase on increasing the water concentration which is consistent 
with the experimental observations qualitatively. In these computer simulations 
\cite{otts3}, the 
stochastic movement of each particle is restricted to their six nearest neighbors 
on a cubic lattice.  Each particle becomes immobile after forming a covalent bond 
with a constant bond length (lattice constant). Although, this approach is a good 
start, the degrees of freedom for particles hopping is minimal and the film
morphology lacks dynamics.

In this paper we introduce a computer simulation model (see below) in which the 
number of degrees of freedom of constituents is increased considerably and the
bond length is allowed to vary. 
The bond fluctuations are important to incorporate the elastic dynamics for 
euilibrating the film morphology. The model is described 
in the next section followed by results and discussion with a summary and 
conclusion at the end.

\section{Model}
We consider three types of particles, hydrophobic ($H$), hydrophilic ($P$), and 
water ($A$) on a simple three dimensional discrete lattice of size $L_x \times
L_y \times L_z$. A particle is described by a cube occupying its eight lattice 
sites. Initially, particles $H,\  P$, and $A$ are randomly distributed on fraction 
$p_H, p_P$, and $p_A$ of the lattice sites with excluded volume constraint which 
entails that a site cannot be occupied by more than one particle. An isolated 
particle can move to one of its 26 neighboring sites. In addition to excluded 
volume (hardcore) interaction, we consider short range interactions among the 
particles at the neighboring sites within a distance $r = 3$. 
The interaction energy E,
$$E = \sum_{ij}J(i,j), \eqno{(1)}$$
where, $i$ runs over each particle and $j$ runs over its neighboring sites within 
a range $r$. 
$$J (H, H) = J(P, P) =  J(H, A) =  J(P, A) = \epsilon. \eqno{(2)}$$
In this study we use $\epsilon = 0, 1$. There is an attractive interaction between 
each particle and the adsorbing substrate (S) at the bottom $z = 1$,
$$J(H, S) = J(P, S) = J(A, S) = 2. \eqno{(3)}$$
Additionally, each component ascribes a molecular weight ($M_H,\  M_P,\  M_A$) in 
arbitrary unit to incorporate the effect of gravitational precipitation. We choose, 
$M_H = M_P = 1,\  M_A = 0.1$ in this study.  The gravitational potential energy at 
height $z$ from the bottom,
$$E_g = M_{H/P/A} \cdot z. \eqno{(4)}$$
The temperature $T$, measured in arbitrary unit (i.e., in unit of the Boltzmann 
constant $k_B$ and the interaction energy), is used as a parameter to control the 
stochastic motion of the particles. Each particle attempts to move to one of its 
randomly selected adjacent site (26) with the Metropolis algorithm. That is, 
a particle at a site say $k$ and one of its neighboring adjacent sites $l$ are 
selected randomly. If the proposed move does not violet the excluded volume 
constraint, then the total energies $E_k$ and $E_l$ in corresponding configurations 
with particle at the initial site ($k$) and its proposed move to randomly selected 
new site ($l$) are evaluated. The particle is moved from site $k$ to $l$ with 
probability $exp(E_k-E_l)/T$. Periodic boundary conditions are used along the 
transverse ($x, y$) directions while open boundary conditions at the top and the 
bottom. Movements of particles are restricted at/near the impenetrable substrate at 
$z = 1$. The aqueous constituent ($A$) can evaporate, i.e., leave the sample from 
the top if it attempts to do so. However, constituents $H$ and $P$ cannot leave the 
sample, their concentrations are thus conserved. Attempts to move each particle 
once defines unit Monte Carlo step (MCS) time. 

Each constituent attempts to react with one of its neighboring particles or 
substrate sites with a fixed reaction rate (probability) after each MCS time. 
In order to implement reaction kinetics, each reacting unit ($H, P$) is assigned a 
functionality of four (in this study), i.e., it can react up to four neighboring 
functional groups including the substrate. The rates (probabilities) of reactions 
with the substrate, $R_{SH} = R_{SP} = 1$ and among the functional groups 
$R_{HH} = R_{HP} = R_{PP} = 1$. Each hopping attempt is followed by covalent 
bonding among the reacting units. Further, each covalent bond resulting from such 
reaction kinetics is irreversible. However, its bond length can fluctuate between 
2 and $\sqrt{10}$ if the bonded unit attempts to move, as in bond fluctuation model 
of a polymer chain \cite{carmesin}.

It is worth pointing out that we are able to incorporate more realistic features 
in this model than the previous studies for such a multicomponent film growth. 
Not only the degrees of freedom for each constituent to execute its stochastic 
movement has increased (from 6 in previous studies \cite{pandey, otts3} to 26 
here), but also the range 
of interaction among the constituents. The second important feature of this model 
is the continued segmental mobility of the covalently bonded units with fluctuating 
bond length of the grown film which is in contrast to absence of mobility 
altogether in previous studies \cite{pandey, otts3} with fixed bond length. 
The elastic nature of the 
film is thus incorporated via vibrating covalent bonds  an important feature 
(the vibration spectra) monitored in IRIR measurements \cite{otts1}. 
This model is still far 
from a complete description of the laboratory system (i.e., arbitrary reaction rate)
but it is a considerable improvement over the previous studies \cite{pandey, otts3}.

Our simulation involves thermodynamic equilibration via stochastic motion of each 
constituent and segmental moves and kinetic reaction while the aqueous component 
continues to evaporate. Depending on the rate of reaction, the thermal equilibration
can be arrested by covalent bonding.  The rate of reaction and temperature along 
with the concentration of each component and their characteristics play important 
roles in designing the film and controlling its characteristics. In this study, 
we are restricted to a small set of these parameters (see below) to illustrate how 
it works.

\section{Results and Discussion}

Initially, all the component particles, 
polar ($P$), hydrophobic ($H$) and water ($A$) are distributed randomly with their 
number concentrations $P_P = 0.01,\  P_H = 0.01$ and $P_A = 0.03$, respectively. 
Note that the volume concentrations are eight times larger than the number 
concentrations as each particle occupies eight lattice sites of a cube. As time 
progresses, e.g., $t=5, 10$, crosslinking grows and film begins to propagate in 
upward direction from the substrate until most of the reactive components are 
crosslinked ($t=40$). A remarkable feature of this model is the ability to move 
crosslinked particles by including the corresponding bond fluctuations even though 
within a limited range (see preceding section). As a result, the crosslinked film 
may become more compact if subjected to downward pressure or gravity. In our 
computer simulation experiments the film thickness reaches a maximum value as water 
continually evaporates and reaches a steady state in asymptotic time limit as 
discussed in the following.

The film consists of all points that are covalently bonded from the substrate. 
The surface is the locus of all connected points with the maximum height. Thus, 
each substrate point ($i$) has a film height $h_i$. The film thickness $h$ is 
defined as,
$$h = {1 \over N_s} \sum_i h_i, \eqno{(5)}$$
where, the number of substrate points $N_s = L_x \times L_y$.

First we examine the simple case in which only hardcore interaction between 
the particles is considered during equilibration resulting from the 
stochastic motion in film formation, i.e., no interaction among polar,
nonpolar and water particles was involved in calculating energy for their
moves. Figure 1 shows the variation of the mean film thickness $h$ with the 
time steps for a range of temperatures ($T=0.5, 1, 1.5, ... , 5$). 
Three stages of film growth are observed for all temperatures: an 
initial fast growth of the mean surface height followed by a relatively slow 
relaxation (kinetic reaction regime) before approaching a steady state. 
The first stage of the film growth may be expressed as
$$h = A t^{\gamma} + B t^{\delta} + ..., \eqno{(6)}$$
with a leading power law index $\gamma$ and correction term with exponent 
$\delta$ for $\gamma > \delta$; $A$ and $B$ are constants.  
For example, at temperature $T = 0.5$ (indicated by 
dashed lines) there may be two growth rates described by $\gamma \simeq 2.0$ 
followed by $\gamma \simeq 1.6$ (see figure 1). 

The initial growth of the film thickness (for $t \le 30$) is followed by 
polymerization in the intermediate relaxation regime ($t \approx 30 - 10^3$) 
where most of the functional groups are already reacted. However, the bonded 
units continue to execute their stochastic movements allowed by corresponding
bond fluctuations as the film thickness $h$ equilibrates to a saturated value 
$h_s$. Mobility of the constituents and corresponding bond lengths depends 
on the temperature, i.e., higher the temperature, larger is the bond length. 
Accordingly, the equilibrium film thickness $h_s$ depend on temperature. 
Variation of $h_s$ with temperature is presented in the inset (see figure 1) 
which shows that, on increasing the temperature, a fast linear expansion of 
the film thickness is followed by relatively slow expansion at high 
temperatures. Unlike the previous studies \cite{pandey, otts3}, where the film 
thickness (surface 
height) remains constant after the reaction reaches its equilibrium 
saturating most of its functional groups, the mean film thickness further
relaxes to its equilibrium value due to bond fluctuations in the asymptotic
time limit. The mean surface height of the film is larger at higher 
temperature which may be due to more active vibration of bonded particles at 
high temperature. It is also clear from these data (figure 1) that the 
relaxation time required to reach equilibrium varies with temperature; 
films at higher temperatures require longer time to equilibrate.

The interface width $W$ of the film surface is defined as the root mean 
square (RMS) fluctuation of the film height \cite{barabasi, family}, i.e., 
the thickness,
$$W^2 = <h_i^2 > - h^2, \eqno{(7)}$$
$$<h_i^2 > = {1 \over N_s} \sum_i h_i^2. \eqno{(8)}$$
Figure 2 shows the variation of the interface width $W$ with the time step 
at different temperatures corresponding to height variation presented in 
figure 1. Initially, the interface width, $W$, grows with time $t$ with a 
power law,
$$W = A t^{\beta}, \eqno{(9)}$$
with exponent $\beta$ which changes from $\beta \approx 1.0$ to $\beta 
\approx 0.80$ in time at the temperature $T = 0.5$ (see figure 2). 
The interface width then decreases with an overshoot (maxima) before 
stabilizing somewhat with an increasing trend at higher temperatures in the 
intermediate time regime. It finally relaxes to steady state saturation in 
the long (asymptotic) time regime.  The general non-monotonic approach 
before reaching saturation persists at all temperatures with obvious change 
in patterns with the temperature. 

Response of the interface width growth can be explained by the three stage 
development of the film formation similar to evolution of the mean surface 
height, $h$. At early stage, films propagate upward unevenly from the 
substrate resulting in increasing fluctuations in the propagating front. 
Since the reaction propagates from the substrate, nearby functional groups 
have high probability to react. As a result, the interface width, i.e., 
the surface roughness increases with time. There after, it decreases as the 
film formation propagates to the top where most particles near the film air 
interface have reacted and become part of the film. The film starts shrinking
toward the substrate as the interstitial water evaporates; the consolidation 
leads to decrease in the surface roughness. The film roughness grows up 
again due to the stochastic movement of the crosslinked particles and 
increasing pore space (empty sites) provided by water evaporation. 
As water content in the film reduces, the film stretches too far with an 
almost second maximum (overshoot) in the interface width (at high 
temperatures). The bond fluctuations respond and the film roughness 
eventually reaches a steady state where both the mean surface height and 
the root mean square interface width remain constant.  A somewhat oscillatory
response, more pronounce with the fluctuation in the film height, i.e., the 
interface width, is due to bond fluctuations of covalent bonded functional 
groups augmented by the interplay between the thermal energy and the free 
volume due to water evaporation. Such a pronounced non-monotonic and 
somewhat oscillatory response of the interface width (roughness) during the
film formation and equilibration is unique characteristics cannot be captured
by constant bond length in previous studies. 

The saturated interface width $W_s$ is found to increase linearly with the 
temperature as shown in the inset figure 2. Similar to the film thickness 
(figure 1 inset), the saturated interface width, $W_s$, shows two linear 
scaling, a fast growth at low temperatures while a relatively slow growth at 
high temperatures. This is in contrast to previous studies \cite{otts3} where
the interface continues to decay with the temperature.

Now let us examine what happens on including more interaction among mobile 
functional groups, i.e., the interaction eq. 1-4. As show in Figure 3, 
the variation of the mean surface height of the film, $h$, with time remains
nearly the same as the system with the hardcore interaction alone (see
fig. 1) at higher temperature. The film thickness $h$ at low temperature 
($T=0.5$) exhibits much slower decrease after reaching its maximum. 
It would be interesting to have a closer look by examining the evaporating
water concentration in the sample. Figure 4 shows the decay of water
concentration as it evaporates during the film growth. Evidently, 
a considerable amount of water still persists in the film at low temperature
($T=0.5$). This provides an insight into the competition between the 
interaction energy and thermal energy causing water to evaporate.
At low temperature, the mobility of the particles is relatively low.
Attractive interaction between water constituents and the the polar groups 
makes it harder for water to escape the lattice which results in a swollen 
film. In this case (low temperature), the film has not reached a steady state
equilibrium. At higher temperature, the overall mobility of particles 
increases, water evaporation is less hindered by its interaction with polar 
particles. Figure 3 (inset) shows this variation of mean surface height 
with temperature which shows a sharp decrease of film thickness at 
low temperatures followed by nearly constant with a slight increase of film 
thickness at high temperatures. Note the sharp contrast between the 
non-monotonic variation of the film thickness here with that of the hardcore
constituents where the film thickness continues to increase with the
temperature. 

Corresponding growth of the interface width $W$ for the interacting 
functional groups is presented in figure 5 for a range of temperatures.
At a first glance, the interface growth pattern for interacting functional 
groups (fig. 5) appears similar to that with the hardcore interaction alone
(fig. 2). However, a closer examination reveals important differences.
The interface width $W$ growth faster for interacting system with a higher
growth exponent. For example, $\beta_1 \simeq 1.15 \pm 0.02$ (initially)
and $\beta_2 \simeq 0.83 \pm 0.02$ (later stage) at $T =0.5$ (fig. 5) in 
comparison to corresponding values $\beta_1 \simeq 1.10 \pm 0.02$ and 
$\beta_2 \simeq 0.81 \pm 0.02$ for hardcore constituents (fig. 2).
The saturated interface width, i.e., the steady state roughness $W_s$ 
decreases sharply with the temperature at low temperatures followed by a 
very slow linear growth on increasing the temperature (inset fig. 5).
This must be contrasted from the linear increase of $W_s$ with temperature
for systems with hardcore interaction alone (inset fig. 2) and that of the
monotonic decay in previous study \cite{otts3}. The decay 
pattern of $W_s$ with temperature for both film thickness ($h$) and the 
interface width ($W_s$) fits relatively well with exponential as well as
power laws; the range of temperature is too small to differentiate between
these empirical trends. 

To make sure that there is no severe finite size effects, simulations are
performed with different lattice sizes. Variations of film thickness and the 
interface width for these lattices are presented in figures 6 and 7 respectively.
There is very little effect of the lattice size on the qualitative nature of the
dependence of film thickness and its interface width. 

\section{Conclusions}

A computer simulation model is proposed to study the film growth and surface 
roughness in a multi-component system consisting of hydrophobic and polar groups
in evaporating aqueous solvent. The thermodynamic equilibration is implemented
by moving these constituents stochastically with the Metropolis algorithm while
incorporating the kinetic reactions from an adsorbing substrate resulting in
flexible covalent bonding. The model is considerable improved over the recent
studies on modeling the film formation in a waterborne polyurethane film while
maintaining the efficiency of a discrete lattice. The degrees of freedom for each 
constituent to move and better relax effectively is much larger, 26 in this study
in comparison to 6 in previous studies. In contrast to fixed bond length and 
immobile bonded units in previous studies, the flexibility of fluctuating 
covalent bonds between the reacted functional groups with ability to perform 
their stochastic movement not only make it a more realistic film growth
but also adds elastic nature of the film.

The simulation is performed for a range of temperatures. Simulations show how
the film thickness grows and its interface evolves and equilibrates. The growth
of the film thickness and surface roughness are found to exhibit power-laws,
i.e., $h \propto t^{\gamma}, \ W \propto t^{\beta}$ with $\gamma \simeq 1.5-2$
and $\beta \simeq 0.8 - 1.0$ in late and early stages of the growth.
Both, the film growth and the interface width show a non-monotonic response while
water evaporates and they equilibrate and approach their steady-state values ($h_s,
\ W_s$). Scaling of these quantities with the temperature shows the effect of
interactions among the constituents. For the components with the hard-core 
interaction alone, the film thickness and the roughness seems to increase linearly
with the temperature in two steps. Including a short range interaction shows that
a sharp decay of both the film thickness and the roughness is followed by a slow
increase on increasing the temperature leading to a non-monotonic dependence.
The non-monotonic dependence of the roughness with temperature is very different
from the monotonic decay pattern of previous study \cite{otts3}. It is somewhat
similar to temperature dependence of roughness in a deposition of polymer chain
model \cite{bentrem}.
Many realistic features and variables are still to be incorporated such as 
rate of reactions, variations in concentration of each component, 
homo- and radical initiated polymerization, etc. some which will be considered
in future in our on-going effort to understand the laboratory experiments. 
 
\bigskip
\bigskip

\noindent
{\bf Acknowledgments:}
Authors thank Marek W. Urban for discussion. 
Support from the National Science Foundation Material Research Science and 
Engineering Center (MRSEC) Program (DMR 023883) and NSF-EPSCoR in part is
acknowledged.


\begin{figure}[hbt]
\begin{center}
\includegraphics[angle=-90,trim=10 0 0 100,width=6.5in]{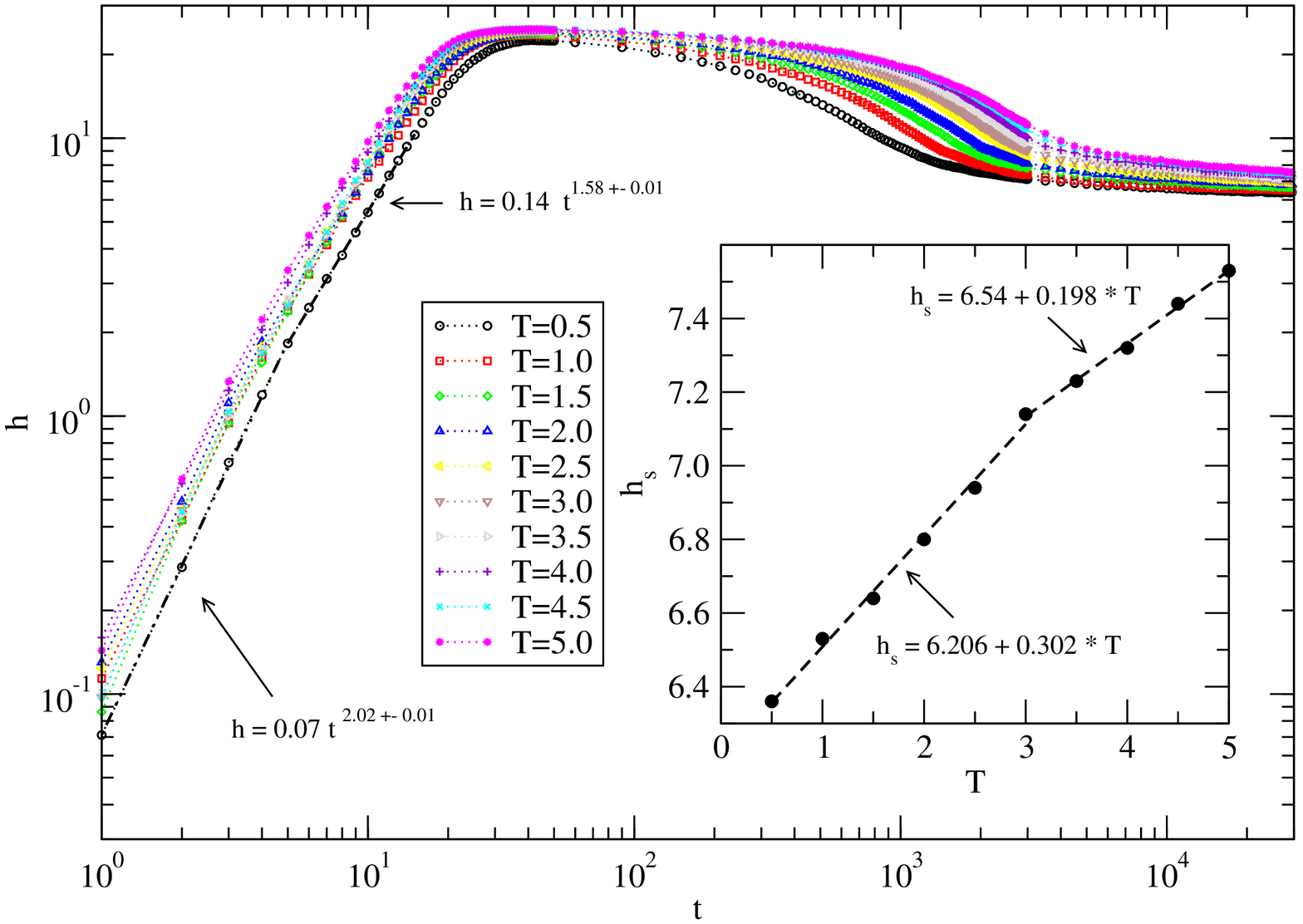}
\end{center}
\caption{
Growth of the average film thickness ($h$) for a range of temperatures for 
hardcore particles on a sample $40 \times 40 \times 30$ with 10 independent
samples for $p_H = p_P = 0.01$ with initial water concentration $p_A = 0.03$.  
The inset figure shows variation of the saturated thickness $h_s$ with the 
temperature on a loglog scale. 
}
\end{figure}

\begin{figure}[hbt]
\begin{center}
\includegraphics[angle=-90,trim=10 0 0 100,width=6.5in]{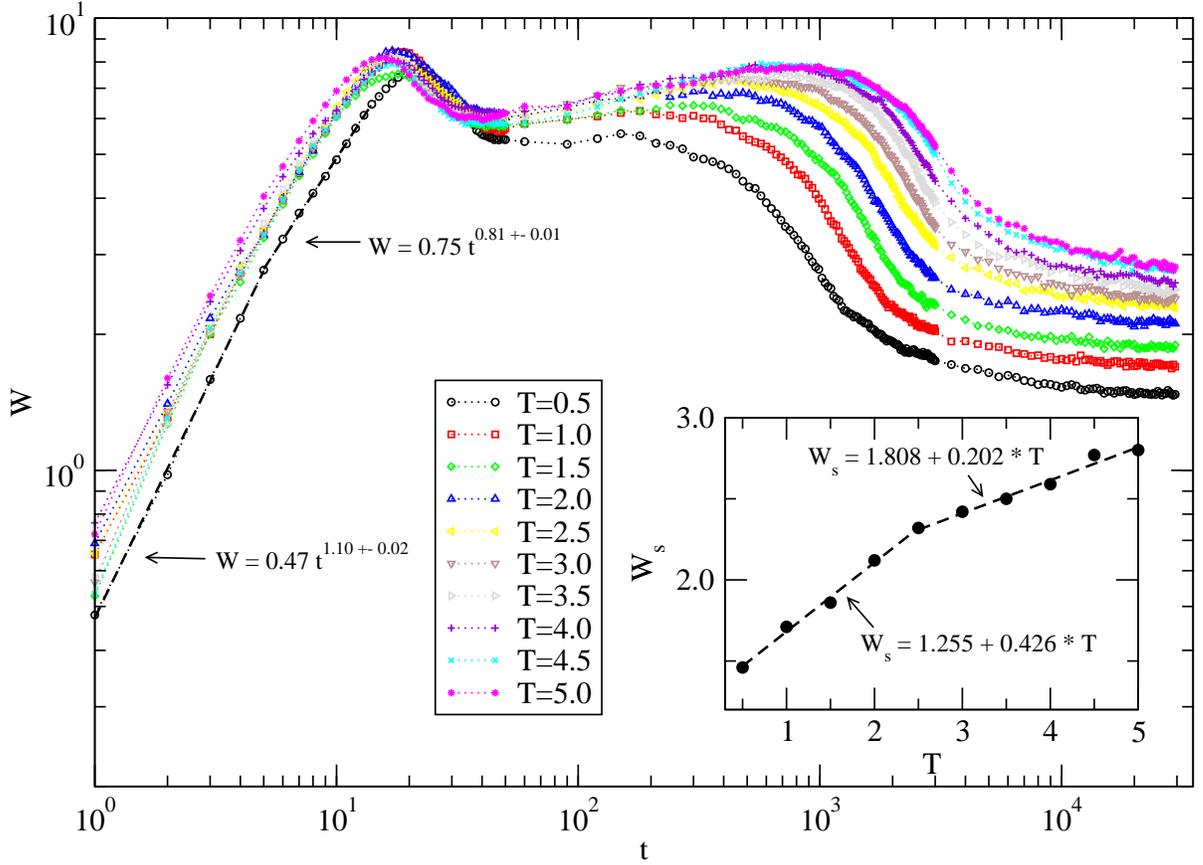}
\end{center}
\caption{
Evolution of the interface width ($W$) of the film surface for a range of 
temperatures for hardcore particles on a sample $40 \times 40 \times 30$ with 10 
independent samples for $p_H = p_P = 0.01$ with initial water concentration 
$p_A = 0.03$. The inset figure shows variation of the saturated interface width,
i.e., the roughness $W_s$ with the temperature. 
}
\end{figure}

\begin{figure}[hbt]
\begin{center}
\includegraphics[angle=-90,trim=10 0 0 100,width=6.5in]{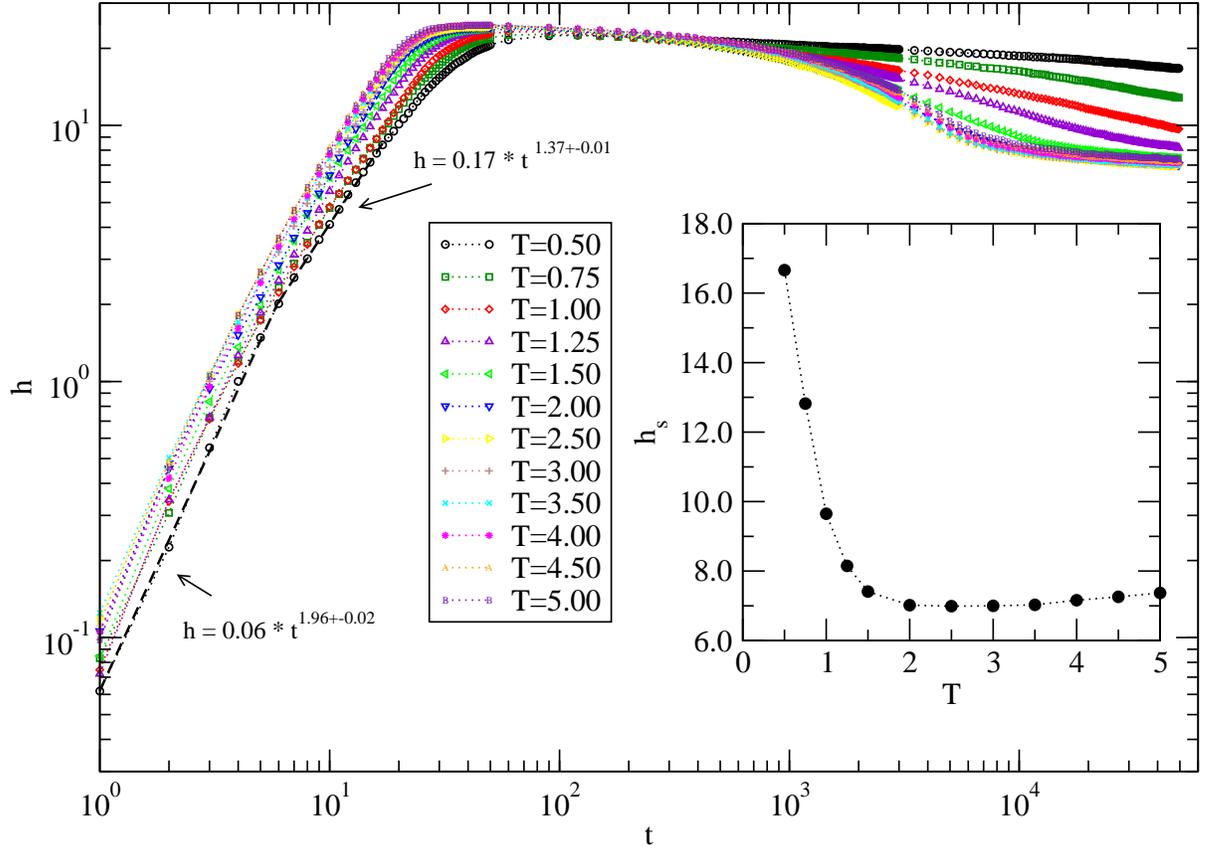}
\end{center}
\caption{
Growth of the average film thickness ($h$) for a range of temperatures for 
interacting particles (eq. 14) on a sample $40 \times 40 \times 30$ with 10 
independent samples for $p_H = p_P = 0.01$ with initial water concentration 
$p_A = 0.03$. The inset figure shows variation of the saturated thickness $h_s$ 
with the temperature. 
}
\end{figure}

\begin{figure}[hbt]
\begin{center}
\includegraphics[angle=-90,trim=10 0 0 100,width=6.5in]{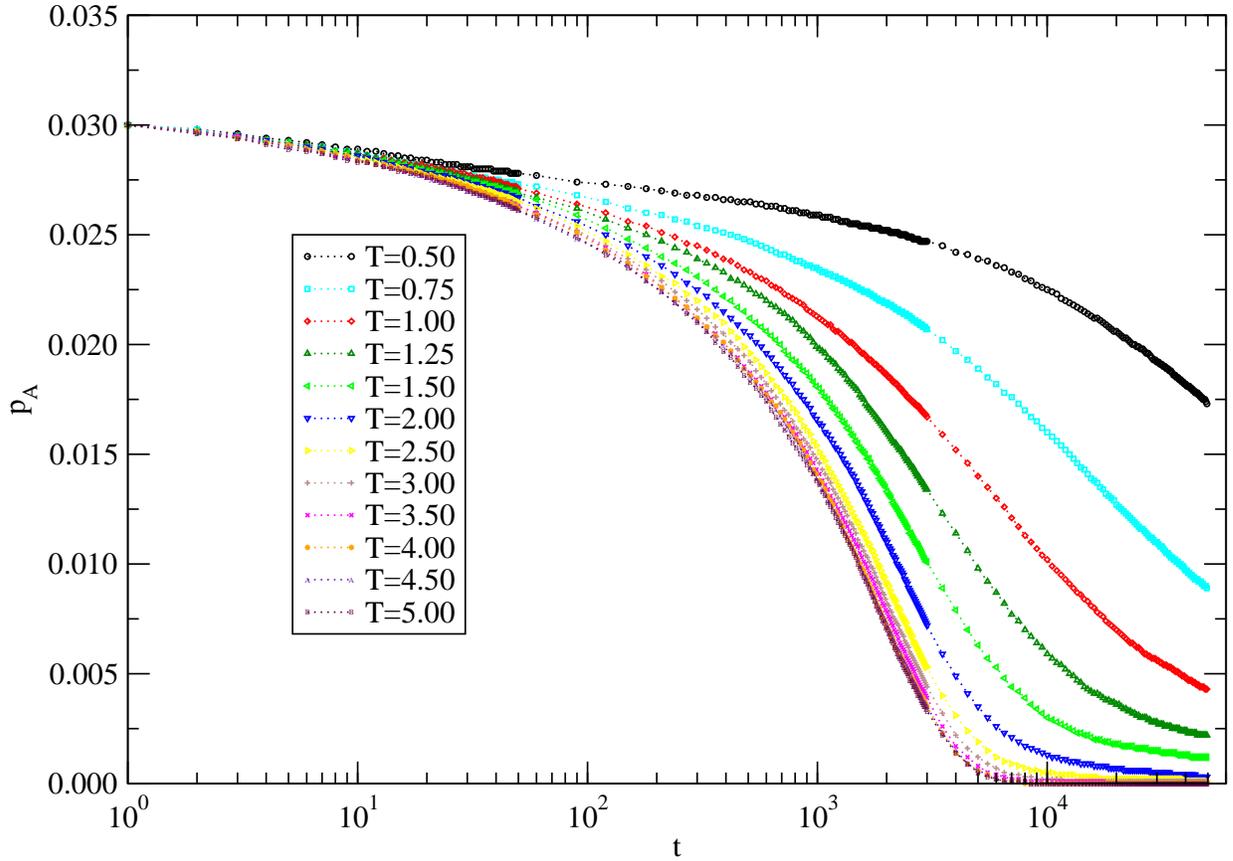}
\end{center}
\caption{
Decay of the evaporating water concentration for a range of temperatures for 
interacting particles (eq. 14) on a sample $40 \times 40 \times 30$ with 10 
independent samples for $p_H = p_P = 0.01$ with initial water concentration 
$p_A = 0.03$. 
}
\end{figure}
 
\begin{figure}[hbt]
\begin{center}
\includegraphics[angle=-90,trim=10 0 0 100,width=6.5in]{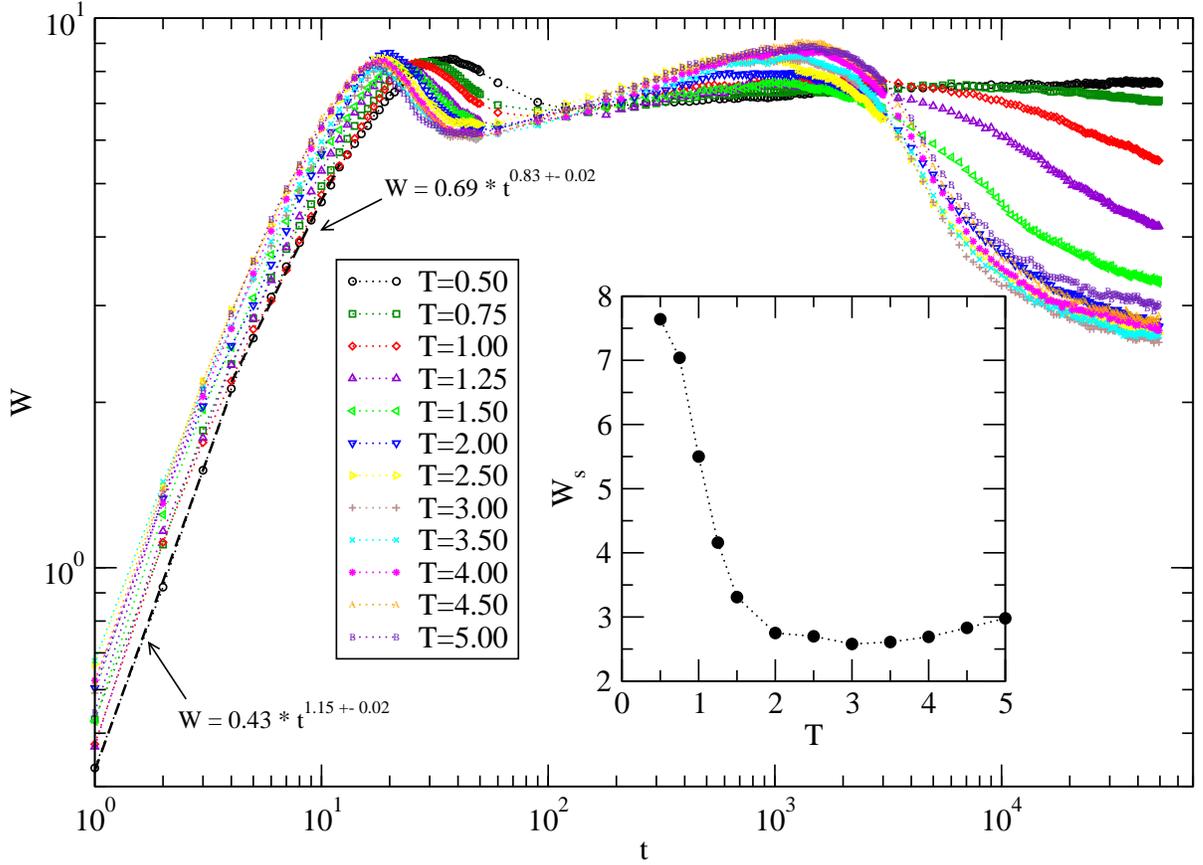}
\end{center}
\caption{
Evolution of the interface width ($W$) for a range of temperatures for 
interacting particles (eq. 14) on a sample $40 \times 40 \times 30$ with 10 
independent samples for $p_H = p_P = 0.01$ with initial water concentration 
$p_A = 0.03$. The inset figure shows variation of the saturated width, i.e.,
the roughness $W_s$ with the temperature. 
}
\end{figure}

\begin{figure}[hbt]
\begin{center}
\includegraphics[angle=-90,trim=10 0 0 100,width=6.5in]{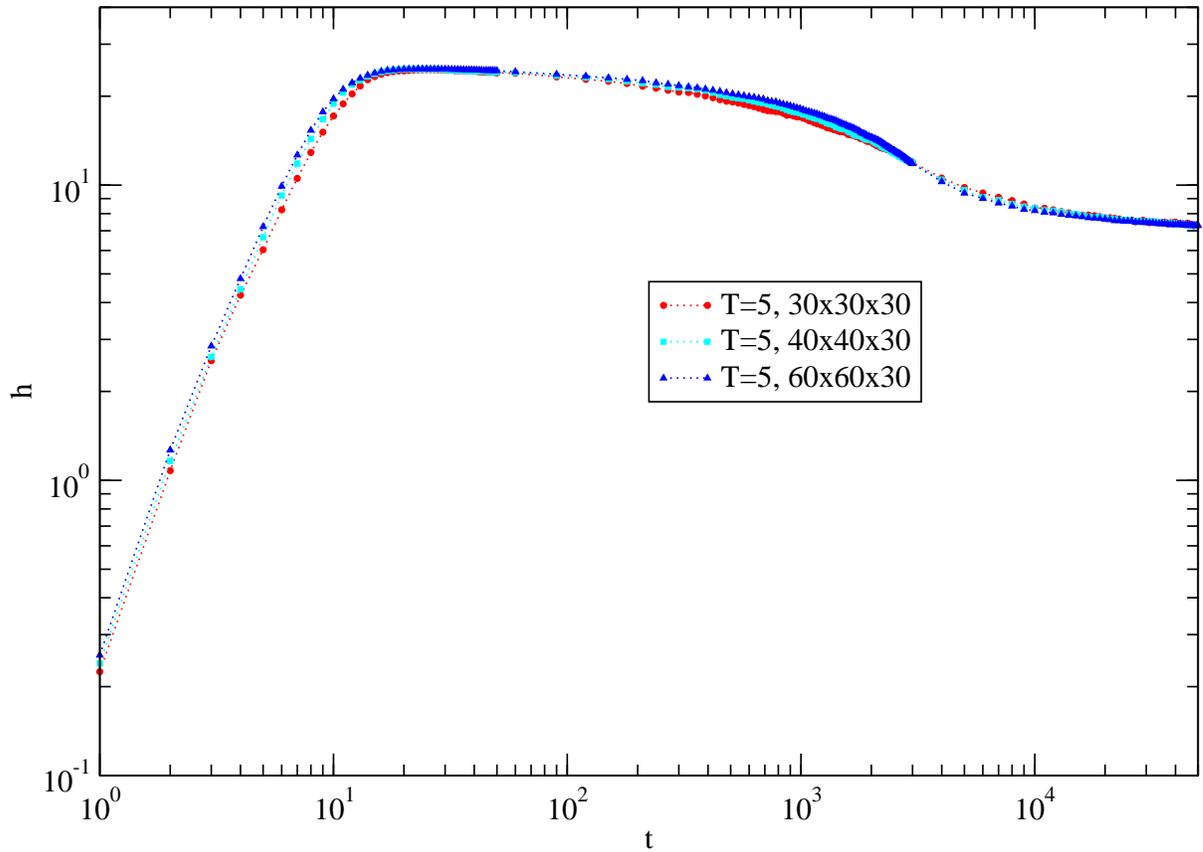}
\end{center}
\caption{
Growth of the average film thickness ($h$) at a temperature $T = 5$ for 
interacting particles (eq. 14) on different sample sizes each with 10 
independent runs for $p_H = p_P = 0.01$ with initial water concentration 
$p_A = 0.03$.
}
\end{figure}

\begin{figure}[hbt]
\begin{center}
\includegraphics[angle=-90,trim=10 0 0 100,width=6.5in]{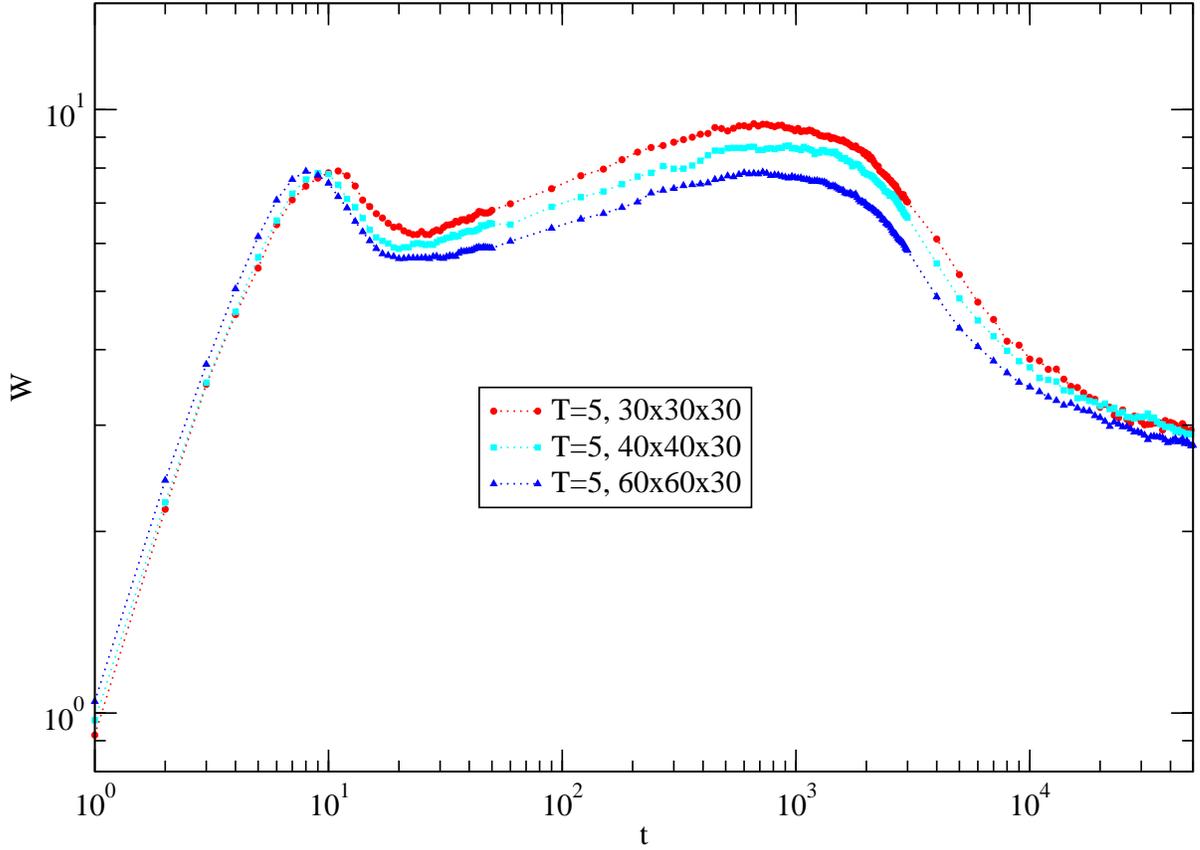}
\end{center}
\caption{
Evolution of the interface width ($W$) for a temperature $T = 5$ for 
interacting particles (eq. 14) on different sample sizes each with 10 
independent samples for $p_H = p_P = 0.01$ with initial water concentration 
$p_A = 0.03$. 
}
\end{figure}

\end{document}